\RequirePackage[2020-02-02]{latexrelease}
\documentclass[aps,prl,twocolumn,superscriptaddress]{revtex4-2}

\usepackage{amsmath}
\usepackage{graphicx}
\usepackage{color}

\newcommand{\YZ}[1]{\textcolor{black}{#1}}

\begin{document}

\title{\YZ{Benign saturation of ideal ballooning instability in a high-performance stellarator}}

\author{Yao Zhou}
\email{yao.zhou@sjtu.edu.cn}
\affiliation{School of Physics and Astronomy, Institute of Natural Sciences, and MOE-LSC, Shanghai Jiao Tong University, Shanghai 200240, China}
\author{K.~Aleynikova}
\affiliation{Max-Planck-Institut f\"ur Plasmaphysik, 17491 Greifswald, Germany}
\author{Chang Liu}
\affiliation{Princeton Plasma Physics Laboratory, Princeton, New Jersey 08543, USA}
\author{N.~M.~Ferraro}
\affiliation{Princeton Plasma Physics Laboratory, Princeton, New Jersey 08543, USA}

\date{\today}

\begin{abstract}
To examine the robustness of the designed 5\% $\beta$-limit for high-performance operation in the W7-X stellarator, we undertake nonlinear magnetohydrodynamic (MHD) simulations of pressure-driven instabilities using the M3D-$C^1$ code. 
Consistent with linear analyses, ideal ballooning instabilities occur as $\beta$ exceeds 5\% in the standard configuration. 
Nonetheless, the modes saturate nonlinearly at relatively low levels without triggering large-scale crashes, even though confinement degradation worsens as $\beta$ increases.
In contrast, in an alternative configuration with nearly zero magnetic shear, ideal interchange modes induce a pressure crash at $\beta= 1\%$.
These results suggest that the standard W7-X configuration might have a soft $\beta$-limit that is fairly immune to major MHD events, \YZ{which enhances the appeal of the stellarator approach to steady-state fusion reactors.}
\end{abstract}

\maketitle

\textit{Introduction.}---To realize magnetic fusion energy, the stellarator concept offers unique advantages for avoiding issues associated with the maintenance and stability of the plasma current \cite{Boozer2005}. 
Recent experimental findings of the advanced W7-X stellarator have successfully demonstrated that optimized magnetic configurations can significantly improve plasma confinement, achieving record-breaking stellarator fusion-triple-product values with modest heating power \cite{Beidler2021}. 
To reach higher performance as more heating power becomes available, it is imperative to ensure magnetohydrodynamic (MHD) stability at high $\beta$ (which denotes the \textit{volume-averaged} ratio of plasma to magnetic pressure throughout this letter).

The $\beta$-limits in stellarator designs are typically set by pressure-driven MHD instabilities \cite{Shafranov1983,Nuhrenberg1993}. 
Originally, W7-X was optimized to be stable against resistive interchange and ideal ballooning modes up to $\beta\approx 5\%$ \cite{Beidler1990,Grieger1992} based on linear stability analyses with limited resolution. 
Since then, numerical tools and computing power have significantly improved. 
Moreover, stellarator plasmas are often found to be nonlinearly stable when driven beyond linear stability thresholds in experiments, with unstable modes saturating at harmlessly low levels \cite{Weller2003,Watanabe2005,Weller2006,Sakakibara2008,DeAguilera2015}. 
This implies that linear stability criteria can be overly restrictive, and the $\beta$-limits they predict may not be hard ceilings.
Hence, it is critical to re-examine the MHD stability of high-$\beta$ W7-X plasmas, nonlinearly in particular.

Previous nonlinear MHD modelings of pressure-driven instabilities in stellarator plasmas have mainly focused on the LHD device \cite{Miura2001,Mizuguchi2009,Miura2010,Ichiguchi2011,Miura2017,Sato2017,Ichiguchi2021,Sato2021}.
{In the inward shifted configuration, LHD can often operate above the Mercier criterion \cite{Watanabe2005,Sakakibara2008}.
% up to a point where low-$n$ interchange modes induce core collapse events \cite{Sakakibara2015} . 
The most recent finding is that the precession of trapped thermal ions in hybrid kinetic-MHD simulations can mitigate low-$n$ interchange modes that would otherwise induce crashes in  MHD modeling \cite{Sato2017,Sato2021} ($n$ denotes toroidal mode number). 
%The latter often feature low-$n$ interchange modes in the core, which are sometimes preceded by high-$n$ ballooning modes in the edge \cite{Sato2017}.
In the outward shifted configuration, the $\beta$-limiting core density collapse (CDC) events \cite{Ohdachi2010} are believed to result from high-$n$ ideal ballooning modes in the periphery \cite{Mizuguchi2009,Ohdachi2017}.
\YZ{It is unclear how applicable the findings of LHD, a conventional heliotron, are to W7-X and other configurations in general.}
There have been simulations of pressure-driven instabilities in low-$\beta$ W7-X plasmas with prescribed current profiles that significantly reduce the peripheral magnetic shear \cite{Suzuki2021}.}
Recently, modeling of soft $\beta$-limits in W7-AS experiments with saturation of low-$n$ resistive instabilities using the JOREK code has also been reported \cite{Ramasamy2024}. 

In this letter, we present the first nonlinear MHD simulations of pressure-driven instabilities in high-$\beta$ W7-X plasmas. 
This is enabled by the stellarator extension of the M3D-$C^1$ code \cite{Zhou2021}, which realizes transport-timescale modeling in complex geometry. 
In the standard configuration, our results agree with the design analyses that ideal ballooning modes become linearly unstable when $\beta$ exceeds 5\%. 
However, the simulations predict nonlinear saturation at benign levels with mild confinement degradation, which is much less severe than the original expectation of losing the plasma column \cite{Beidler1990} {or the CDC events in LHD \cite{Ohdachi2010,Ohdachi2017}}. 
This suggests that the designed W7-X $\beta$-limit might be soft, but by no means implies that nonlinear stability is guaranteed in stellarators.
The latter point is exemplified in an alternative configuration with nearly zero magnetic shear, in which ideal interchange modes trigger a pressure crash at $\beta= 1\%$.
Therefore, linear MHD stability should still be treated seriously in stellarator design, \YZ{to which nonlinear considerations can be complementary by rewarding some extra cushion. 
This favorable feature strengthens the stellarator's advantage in realizing stable fusion reactors.}

\textit{Simulation settings.---}M3D-$C^1$ is a sophisticated nonlinear MHD code designed to model the macroscopic dynamics of toroidal fusion plasmas \cite{Jardin2012}. 
%For temporal discretization, M3D-$C^1$ implements a split-implicit scheme that realizes stable and efficient transport-timescale simulations \cite{Jardin2012b}. 
%For spatial discretization, M3D-$C^1$ employs high-order finite elements with $C^1$ continuity in three dimensions and has successfully been extended to stellarator geometry \cite{Zhou2021}. 
Its stellarator extension \cite{Zhou2021} has recently been validated by simulations of current-drive-induced sawtooth-like crashes in W7-X, which show semi-quantitative agreement with experimental results \cite{Zhou2023}.
%Recent simulations of current-drive-induced sawtooth-like crashes in W7-X validates this new capability by showing semi-quantitative agreements with experimental results \cite{Zhou2023}.
%While two-fluid and many other effects are available in M3D-$C^1$, their functionality has not yet been fully verified in stellarator geometry. 
In this work, we solve the single-fluid extended MHD equations (in SI units)
\begin{align}
\partial_t \rho + \nabla\cdot(\rho\mathbf{v}) &= D\nabla^2(\rho-\rho_0),\label{continuity}\\
\rho(\partial_t \mathbf{v} + \mathbf{v}\cdot\nabla\mathbf{v}) &= \mathbf{j}\times\mathbf{B} - \nabla p - \nabla\cdot\mathbf{\Pi},\label{momentum}\\
\partial_t p + \mathbf{v}\cdot\nabla p +\gamma p\nabla\cdot\mathbf{v} &= (\gamma-1)[\eta ({j}^2-{j}_0^2)\nonumber\\
&~~~~ - \nabla\cdot\mathbf{q} - \Pi : \nabla\mathbf{v}],\label{energy}\\
%\partial_t p/(\gamma-1)  &\sim \eta ({j}^2-{j}_0^2)\nonumber\\
%\partial_t (B^2/2)  &\sim - \eta \mathbf{j}\cdot(\mathbf{j}-\mathbf{j}_0)\nonumber\\
\partial_t \mathbf{B} &=  \nabla\times[\mathbf{v}\times\mathbf{B} - \eta (\mathbf{j}-\mathbf{j}_0)],\label{induction}
\end{align}
for the mass density $\rho$, velocity $\mathbf{v}$, pressure $p$, and magnetic field $\mathbf{B}$, with the current density $ \mathbf{j}= \mu_0^{-1}\nabla\times\mathbf{B}$ and the adiabatic index $\gamma=5/3$. 
The viscous stress tensor is $\mathbf{\Pi} = -\mu(\nabla\mathbf{v}+\nabla\mathbf{v}^{\text{T}}) - 2(\mu_{\text{c}}-\mu)(\nabla\cdot\mathbf{v})\mathbf{I}$ and the heat flux 
%\mathbf{q} = -(\kappa_\perp\nabla + \kappa_\parallel\mathbf{b}\mathbf{b}\cdot\nabla)(T-T_0)
$\mathbf{q} = -\kappa_\perp\nabla(T-T_0) - \kappa_\parallel\mathbf{b}\mathbf{b}\cdot\nabla T$
%\delta\mathbf{q} = \kappa_\parallel\mathbf{b}\mathbf{b}\cdot\nabla T_0
with $\mathbf{b} =\mathbf{B}/B$ and the temperature $T=Mp/\rho$, where $M$ is the ion mass. 

The transport coefficients used include uniform mass diffusivity $D=2.18~\mathrm{m^2/s}$, isotropic and compressible viscosities $\mu=\mu_\text{c}=3.65\times10^{-7}~\mathrm{kg/(m\cdot s)}$, and strongly anisotropic parallel and perpendicular thermal conductivities $\kappa_{\parallel}=10^{6}\kappa_{\perp}$ with $\kappa_{\perp}=2.18\times10^{20}~\mathrm{(m\cdot s)^{-1}}$. 
The temperature-dependent resistivity $\eta$ is enhanced by 100 times from the Spitzer resistivity and its typical value is about $10^{-6}~\mathrm{\Omega\cdot m}$ (core) to $10^{-5}~\mathrm{\Omega\cdot m}$ (edge). 
The equilibrium fields $\rho_0$, $T_0$, and $j_0$ subtracted in the dissipative terms act as effective sources to sustain the equilibrium in the absence of instabilities.
The equilibrium density and temperature profiles are $\rho_0\propto p_0^{1/\gamma}$ and $T_0\propto p_0^{1-1/\gamma}$.
We use 3807 reduced quintic elements in the $(R,Z)$ plane and 160 Hermite cubic elements in the toroidal direction.
Unless otherwise noted, we use a time step size of 0.573 $\mu$s to simulate a hydrogen plasma with a core number density of $1.5\times10^{20}~\mathrm{m}^{-3}$, such that the core temperature is about 5 keV (assuming equipartition). {A typical simulation consumes about 400,000 CPU hours}.

We initialize fixed-boundary stellarator simulations in M3D-$C^1$ using the outputs of the widely used 3D equilibrium code VMEC \cite{Hirshman1983}, including the geometry of the flux surfaces as well as the magnetic field $\mathbf{B}_0$ and the pressure $p_0$.
The former is given in terms of a coordinate mapping, $R(s,\theta,\varphi)$ and $Z(s,\theta,\varphi)$ with $s$ being the normalized toroidal flux and $\theta$ the poloidal angle in VMEC, which is utilized to set up the non-axisymmetric computational domain {enclosed by the last closed flux surface}. (This mapping does not evolve in M3D-$C^1$ so that $s$ and $\theta$ are fixed regardless of the actual dynamics of the flux surfaces.)
The latter then provide the initial conditions and subtracted equilibrium fields in the simulations. 
{The boundary conditions are ideal on the magnetic field, no-slip on the velocity, and fixed on the density and pressure.}

%We have tried varying the parameters and obtained qualitatively similar results.

\begin{figure}
\includegraphics[width=\columnwidth]{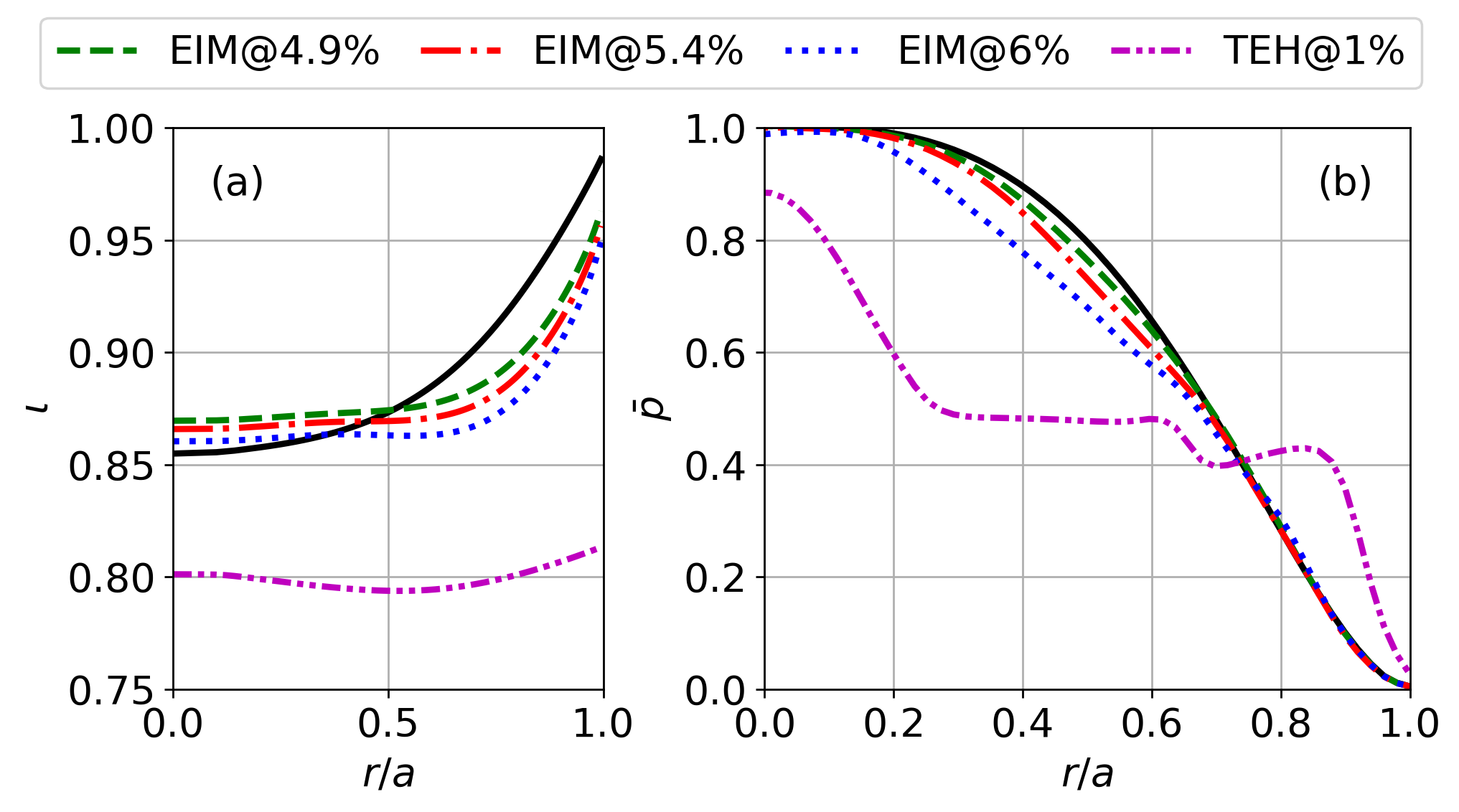}
\caption{(a) {The rotational transform profiles in the simulated VMEC equilibria, with the vacuum profile (EIM, solid) shown for comparison. (b) The final shapes of the pressure profile $\bar{p}$ at $\theta=0$ and $\varphi=\pi$ in the M3D-$C^1$ simulations, with the initial shape (solid) shown for comparison.} Here the normalized minor radius $r/a=\sqrt{s}$.}
\label{profiles}
\end{figure}

\textit{Standard configuration at high $\beta$.---}First, we construct high-$\beta$ free-boundary VMEC equilibria in the standard `EIM' configuration of W7-X. (The three letters in the code reflect the mirror ratio, the rotational transform $\iota$, and the horizontal shift, respectively \cite{Aleynikova2022}.)
The shape of the pressure profile is close to parabolic (the often-used $p\propto 1-s$ profile) but flatter in the core, and the maximum gradient is shifted slightly inward from the boundary in order to mitigate the impact of the fixed boundary. 
(Similar profiles have been used in studies of kinetic ballooning modes \cite{Aleynikova2022,Mulholland2023}.)
The equilibrium $\beta$ is then varied by re-scaling the pressure profile. 
Specifically, we model three cases with $\beta=4.9\%$, $5.4\%$, and $6\%$, respectively, and $B\approx 2.3~\text{T}$ in the core.
The toroidal current profile is prescribed to be zero assuming that the small bootstrap current can be well cancelled by the external current drive \cite{Geiger2015}. 
The rotational transform profiles of these equilibria are shown in Fig.~\ref{profiles}(a). 
Compared with the vacuum profile (solid), the core magnetic shear and edge $\iota$ are lowered as $\beta$ increases, \YZ{but the overall change is quite small. This is an important feature of the quasi-isodynamic concept that W7-X adopts, which minimizes the pressure-driven Pfirsh-Schl\"uter and bootstrap currents and thereby finite-$\beta$ effects \cite{Helander2009,Nuhrenberg2010}}.

In Ref.~\cite{Beidler1990}, local linear stability analyses predict that ideal ballooning modes 
%(with poloidal and toroidal mode numbers $m=8$ and $n=7$, respectively) 
become unstable at $\beta=4.7\%$, and that the resistive interchange threshold is $\beta=7.5\%$. 
In Ref.~\cite{Grieger1992} the $\beta$-limit given is 4.3\%. 
The exact details of how these results are obtained (e.g., the shape of the pressure profile used) are unclear, yet the general perception is that the W7-X $\beta$-limit is around 5\% and imposed by ideal ballooning. 
Consistent with such expectations, the high-$\beta$ EIM equilibria shown in Fig.~\ref{profiles}(a) are Mercier stable but found to be locally ideal ballooning unstable by the COBRA code \cite{Sanchez2000}. This situation is more similar to the outward shifted LHD configuration than inward, \YZ{but it is noteworthy that LHD is subject to more pronounced finite-$\beta$ effects than the quasi-isodynamic W7-X.}

To ensure that the simulations can reasonably model the ideal ballooning modes, we examine how the results depend on toroidal resolution and resistivity using single-field-period (1/5 of the torus) simulations. 
We obtain higher linear growth rates and finer mode structures as the toroidal resolution increases, consistent with the prediction of asymptotic theory that the most unstable mode occurs at $n\rightarrow\infty$ \cite{Connor1979,Dewar1981}. 
That said, as the toroidal resolution increases from 32 to 48 the change in growth rates and mode structures is limited. 
Given the finite computational resources at hand, we choose to use 32 elements per field period (160 in total) in full-torus simulations. 
Meanwhile, as $\eta\rightarrow 0$ the growth rates appear to converge to finite values, which is also consistent with the feature of ideal ballooning.
Coincidentally, since the growth rates increase with $\eta$, the 100-time enhanced Spitzer resistivity somewhat compensates for the reduction in growth rates due to the toroidal resolution.

\begin{figure}
\includegraphics[width=\columnwidth]{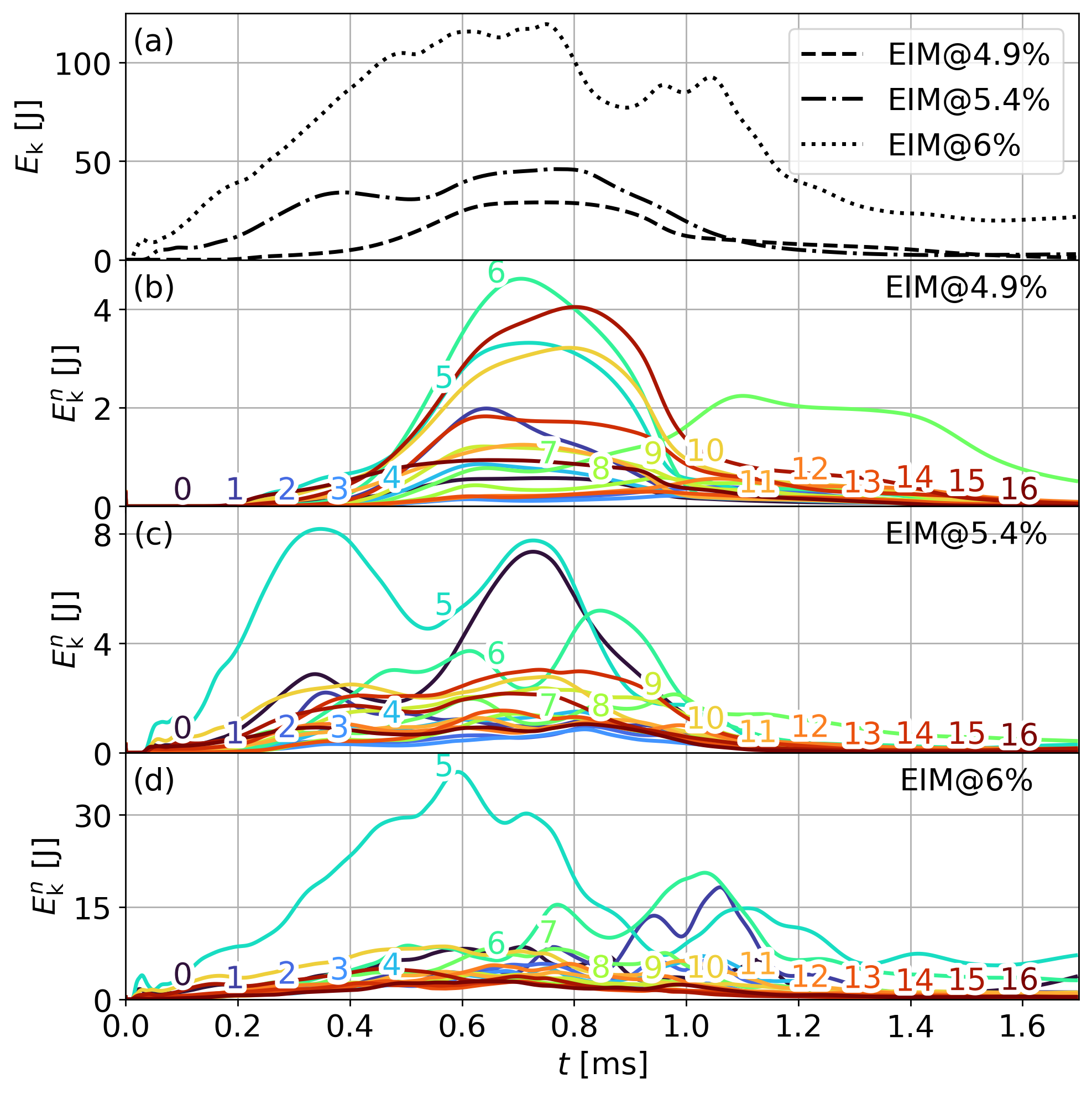}
\caption{The evolution of the total kinetic energy (a) and its toroidal Fourier components (b-d) in the simulations of the three EIM equilibria at high $\beta$, which exhibits limited low-$n$ activities.}
\label{history}
\end{figure}

\begin{figure}
\includegraphics[width=\columnwidth]{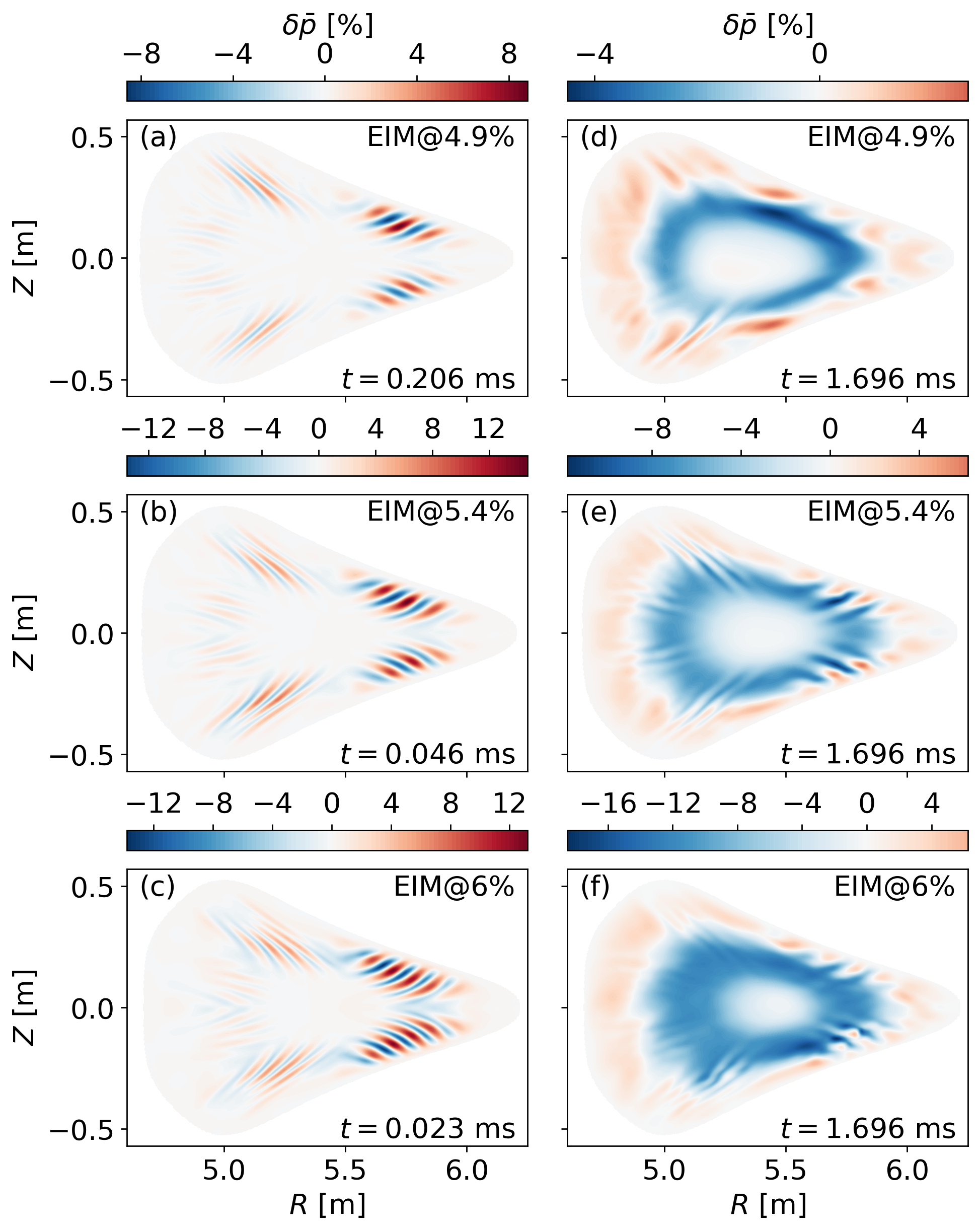}
\caption{Snapshots of the pressure change (in percentage) in the three EIM simulations: (a-c) mode structures at the end of the linear growth phase; (d-f) saturated state at the end of the simulations. Note the different scales in the colorbars.}
\label{EIM}
\end{figure}

Figure \ref{history}(a) shows the history of the total kinetic energy {$E_\text{k} = \int(\rho v^2/2)\,dV$} in the three EIM simulations. 
The linear growth phase is quite short, taking only tens ($\beta=5.4\%$ and $ 6\%$) to hundreds ($\beta=4.9\%$) of microseconds. 
Snapshots of the pressure change $\delta \bar{p} = \bar{p} - \bar{p}_0$ (bar denotes normalization by the initial core pressure) at the end of the linear growth phase are presented in Fig.~\ref{EIM}(a-c), which manifest coherent ballooning mode structures localized in the bad-curvature regions. 
Subsequently, the kinetic energies continue to grow but in a non-explosive manner, then slowly decrease, and eventually saturate at low levels. Animations showing the entire evolutions of $\delta \bar{p}$ can be found in the supplemental material.

Figure \ref{history}(b-d) depicts the history of the toroidal Fourier components of kinetic energy $E_\text{k}^n$ in the three simulations, respectively. 
Note that $E_\text{k}^n$ accounts for the Fourier transform of the velocity only (assuming that the density does not deviate much from the equilibrium) with respect to the toroidal angle $\varphi$, which may not be the most accurate but nevertheless provides useful spectral information on the dynamics. 
In all cases the intrinsic $n=5$ mode family is the most prominent, and as $\beta$ increases the $n=5$ mode becomes more dominant. 
At $\beta=4.9\%$ we also see the $n=6$ and $n=7$ modes grow, and at $\beta=5.4\%$ the $n=0$ mode as well. 
However, there is no significant occurrence of lower-$n$ ($n<5$) activities, {arguably due to the lack of lower-$n$ resonances in the configuration.}

\begin{figure}
\includegraphics[width=\columnwidth]{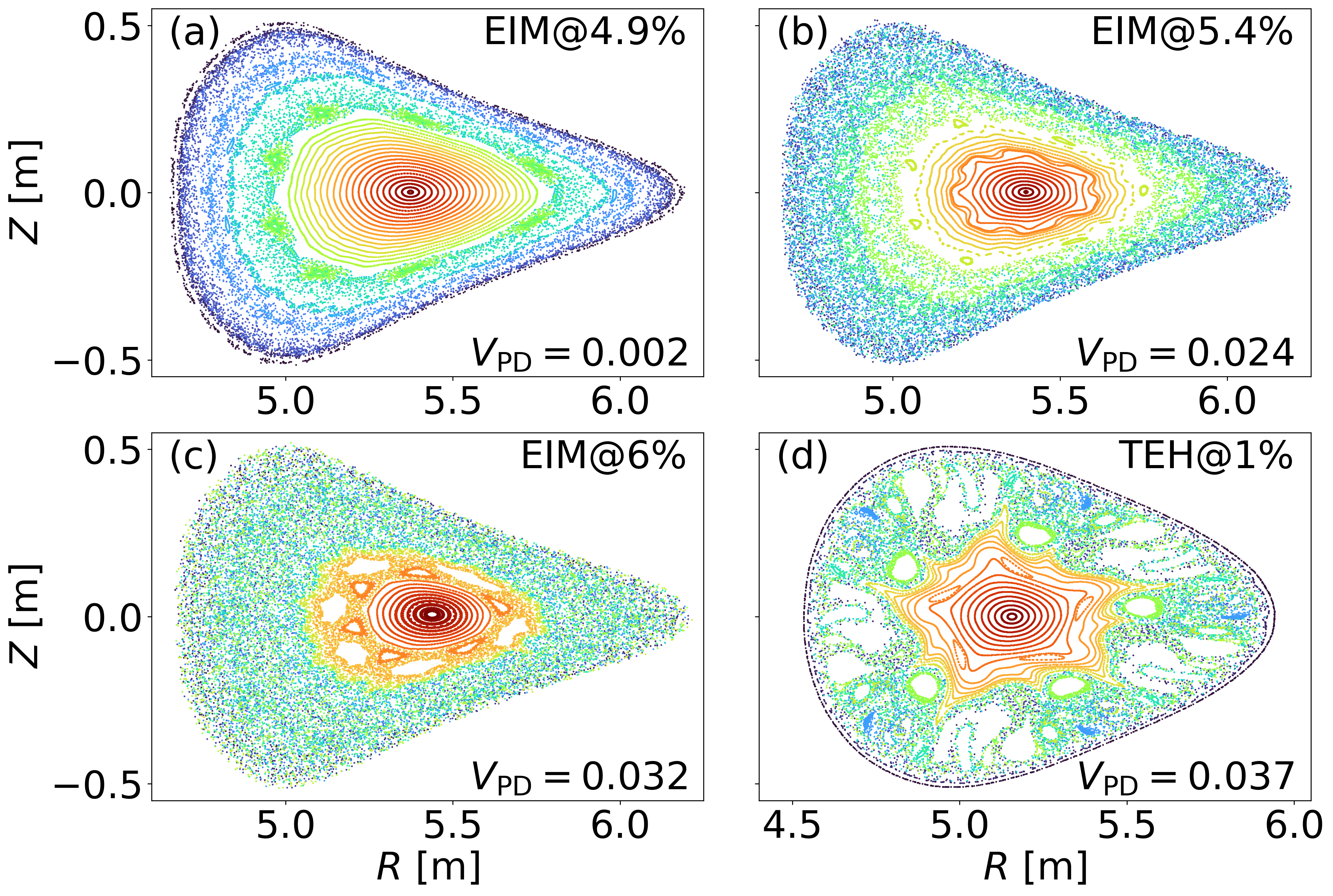}
\caption{Snapshots of Poincar\'e plots obtained at the end of the four simulations: (a-c) correspond to Fig.~\ref{EIM}(d-f), and (d) corresponds to Fig.~\ref{TEH}(d). Colors label different field lines.}
\label{poincare}
\end{figure}

The snapshots of the saturated pressure change in Fig.~\ref{EIM}(d-f) show the profile changes due to nonlinear coupling eventually dominate over the high-$n$ mode structures in Fig.~\ref{EIM}(a-c). 
The corresponding Poincar\'e plots in Fig.~\ref{poincare}(a-c) show that the level of non-integrability in the magnetic field increases with $\beta$. 
Accordingly, in Fig.~\ref{profiles}(b) the change in the shape of pressure profile, which serves as a measure for confinement degradation, also consistently increases.
At $\beta=4.9\%$ the stratified color code reveals copious non-trivial structures, possibly embedded with cantori, which work as transport barriers that reduce temperature flattening \cite{Hudson2008,Hudson2009}. 
The stratification is less pronounced at $\beta=5.4\%$, and almost non-existent at $\beta=6\%$, and therefore confinement degradation is exacerbated.
In particular, the sizable islands with poloidal mode number $m=6$ in Fig.~\ref{poincare}(c) account for the most significant profile change in Fig.~\ref{profiles}(b) (dotted curve) among the EIM cases.

Overall, the changes in pressure profile seem rather small considering how non-integrable the magnetic fields have seemingly become, but this is not uncommon in modelings of anisotropic heat transport in 3D fields \cite{Hudson2009,Helander2022,Paul2022,Suzuki2024} or experiments \cite{Weller2006,Sakakibara2008,Reiman2021}. 
{Following Refs.~\cite{Paul2022,Baillod2023}, we calculate and show in Fig.~\ref{poincare} the effective volume of parallel diffusion $V_\text{PD} = V^{-1}\int\mathcal{H}(\kappa_\parallel|\nabla_\parallel T|^2-\kappa_\perp|\nabla_\perp T|^2)\,dV$, where $V$ is the total volume and $\mathcal{H}$ the Heaviside function. 
Its value increases slightly with $\beta$ but remains small, which indicates that the transport enhancement due to magnetic stochasticization is indeed modest.}

Given that the instabilities are ideal and the originally expected consequence was a destruction of the plasma column like disruptions in tokamaks \cite{Beidler1990}, {or by comparison to the CDC events in LHD \cite{Ohdachi2010,Ohdachi2017},} the limited profile change predicted by our simulations is fairly benign.
This suggests that W7-X may be able to operate at or slightly above $\beta=5\% $ while avoiding major MHD events.
In other words, the standard W7-X configuration may have a soft $\beta$-limit.
{To our knowledge, in stellarator modeling, this is the first report of the benign saturation of: (i) ideal instabilities in pure MHD, in contrast to the kinetic stabilization effects discussed in Ref.~\cite{Sato2021}; and (ii) ballooning instabilities, unlike the pressure crashes they induce in Refs.~\cite{Mizuguchi2009,Miura2010,Miura2017,Sato2017,Suzuki2021}.}

\textit{Alternative unstable configuration.---}The results above confirm that the MHD stability optimization performed in the W7-X design is largely successful, to which the softening of the $\beta$-limit is a bonus.
In fact, had linear stability not been optimized for, severe MHD events could easily occur.
This is exemplified by the following simulation of a highly unstable alternative W7-X configuration.

The magnetic configuration in W7-X can be varied by adjusting the coil currents \cite{Geiger2015}. 
The `TEH' configuration we consider has a higher mirror ratio, lower $\iota$ and outward shift compared with the EIM configuration \cite{Aleynikova2022}. 
We construct a free-boundary VMEC equilibrium using the same pressure and current profiles as above, with $\beta=1\%$ and $B \approx 2~\text{T}$ in the core.
The core number density is $3\times10^{19}~\mathrm{m}^{-3}$ and electron temperature is about 3 keV, and the time step size is halved to 0.286 $\mu$s due to the smaller Alfv\'en time. All other settings are unchanged.

\begin{figure}
\includegraphics[width=\columnwidth]{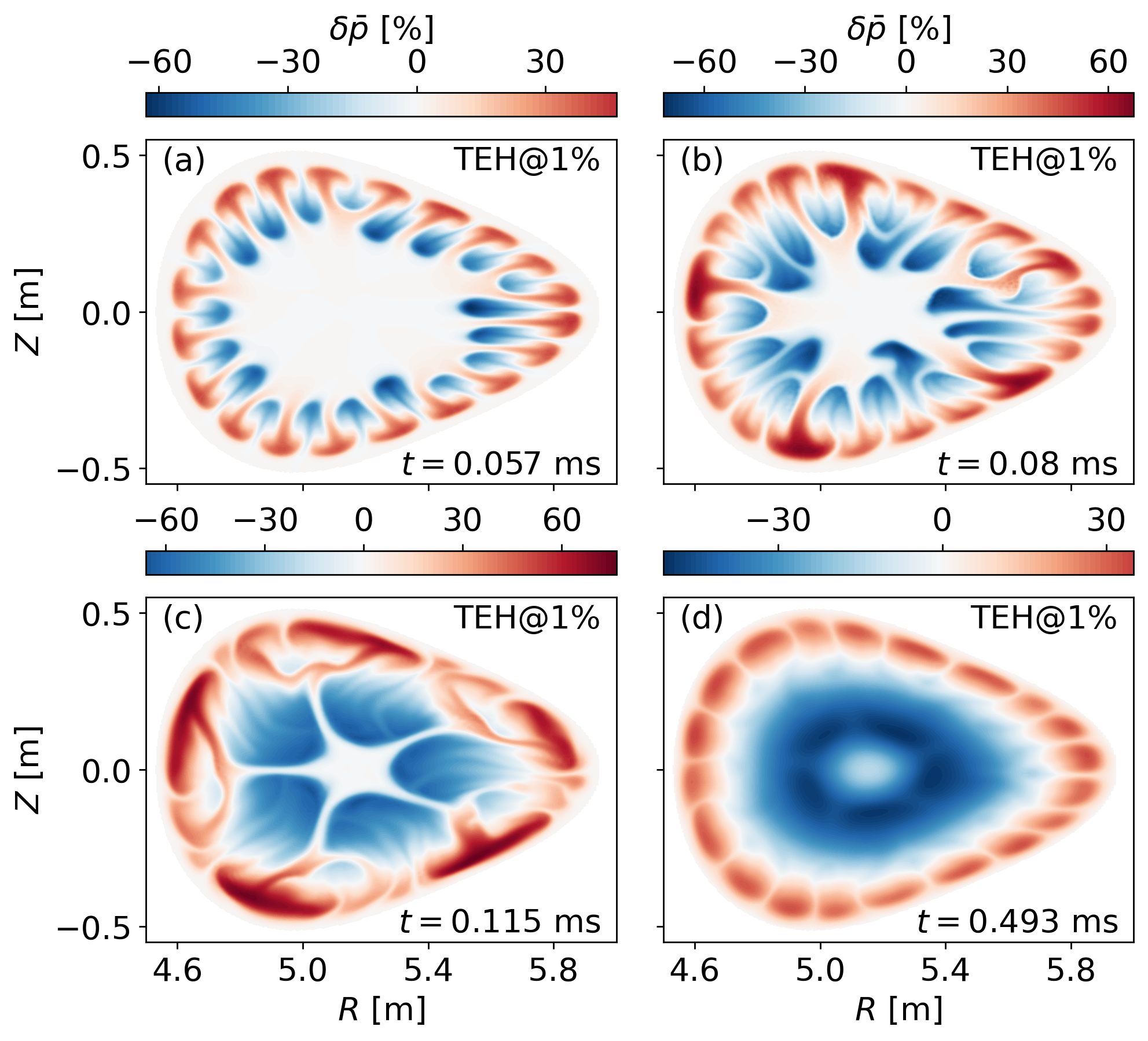}
\caption{Snapshots of the pressure change (in percentage) in the TEH simulation: (a-c) nonlinear evolution of the interchange modes; (d) final post-crash state.}
\label{TEH}
\end{figure}

The $\iota$ and pressure profiles of the TEH equilibrium are displayed in Fig.~\ref{profiles}(a) as well. 
With $\iota\approx 0.8$ and nearly zero magnetic shear, the equilibrium is Mercier unstable.
Indeed, ideal interchange modes quickly develop in the simulation, whose nonlinear evolution is illustrated in Fig.~\ref{TEH}.
In the early phase (a) we see that the fastest-growing mode has $m=20$. 
The cold bubbles then merge into bigger ones with $m=5$ as they propagate from the periphery to the core (b-c), inducing a large pressure crash (d). 
Accordingly, the post-crash pressure profile in Fig.~\ref{profiles}(b) is completely flattened from $r/a\approx 0.25$ to 0.75, and the Poincar\'e plot in Fig.~\ref{poincare}(d) showcases evident remnants of the $m=5$ interchange structure.
{The value of $V_\text{PD}$ is still small, which suggests that the crash is mainly caused by convection rather than conduction.}

By comparison, the interchange instability in the TEH configuration causes significantly more profile change than ballooning in the EIM configuration. 
This implies that low-$n$ interchange instabilities tend to be more detrimental than high-$n$ ballooning and hence should be prioritized in stellarator optimization. 
More importantly, nonlinear MHD stability should not be taken for granted and linear stability is still instrumental in stellarators.  
Nonlinear considerations should not replace linear analyses, but complement them by finding some extra cushion in designs and operations.

\textit{Discussion.---}\YZ{In this work, we use state-of-the-art nonlinear MHD simulations to predict the benign saturation of ideal ballooning modes in the standard W7-X configuration, rendering the designed $\beta$-limit soft. 
For comparison, we also show that interchange modes can induce a pressure crash in an alternative configuration at low $\beta$. 
These results indicate that MHD stability optimization is indispensable yet rewarding in stellarator design, and enhance the stellarator's appeal for reactor applications.}

\YZ{Improvements can be made in several aspects.} 
First, the fixed-boundary treatment may underestimate the confinement degradation due to ballooning modes and also artificially stabilize external modes. 
We can initialize free-boundary M3D-$C^1$ simulations from HINT equilibria \cite{Suzuki2017}, but those seem quite different from the VMEC equilibria at high $\beta$ and need to be further analyzed. 
Second, important physics such as the stabilizing effect of trapped thermal ions \cite{Sato2021} are not yet included. 
A hybrid kinetic-MHD model with thermal ions has been implemented in M3D-$C^1$ \cite{Liu2022} and is being extended to stellarator geometry.
Third, verification against other codes including MIPS \cite{Sato2017,Ichiguchi2021,Sato2021}, JOREK \cite{Ramasamy2024,Nikulsin2022}, and NIMROD \cite{Patil2023}, as well as validation against existing experimental results \cite{Weller2003,Watanabe2005,Weller2006,Sakakibara2008,DeAguilera2015}, are warranted.
All these possibilities will be pursued in future work.

\acknowledgments We thank J.~Geiger, C.~N\"uhrenberg, M.~Sato, and C.~Zhu for helpful comments and discussions, {and the anonymous reviewer for suggesting $V_\text{PD}$ as a metric}. Y.Z. was supported by the National Natural Science Foundation of China under Grant Number 12305246 and the Fundamental Research Funds for the Central Universities.
C.L. and N.M.F. were supported by the U.S. Department of Energy under contract number DE-AC02-09CH11466. 
The United States Government retains a non-exclusive, paid-up, irrevocable, world-wide license to publish or reproduce the published form of this manuscript, or allow others to do so, for United States Government purposes.

\end{document}